\documentclass[11pt]{article}

\usepackage{amssymb}
\usepackage{amsmath}
\usepackage{cite}
\usepackage{graphicx}

\renewcommand{\thefootnote}{\arabic{footnote}}
\newcommand\ZZ{\hbox{\zfont Z\kern-.4emZ}}
\font\zfont = cmss10 

\begin{document}

\begin{titlepage}
\vskip.5cm
\begin{center}
{\huge{ \sc Essay on Gravitation}}
\vskip.3cm
{\huge{\bf The Cosmological Constant Problem}}
\vskip.15cm
{\huge{\bf in Brane--Worlds and }}
\vskip.2cm
{\huge{\bf Gravitational Lorentz Violations}}
\vskip.2cm
\end{center}
\vskip0.2cm

\begin{center}
{\sc Csaba Cs\'aki}$^{a,}$\footnote{J. Robert Oppenheimer
fellow.}, {\sc Joshua Erlich}$^{a}$ and
{\sc Christophe Grojean}$^{b,c}$ \\

\end{center}
\vskip 10pt

\begin{center}
$^{a}$ {\it Theory Division T--8,
Los Alamos National Laboratory, Los Alamos, NM 87545, USA} \\
\vspace*{0.1cm}
$^{b}$ {\it Department of Physics,
University of California, Berkeley, CA 94720, USA} \\ \vspace*{0.1cm}
$^{c}$ {\it Theoretical Physics Group,
Lawrence Berkeley National Laboratory, \\ Berkeley, CA 94720, USA} \\
{\tt email addresses:} csaki@lanl.gov, erlich@lanl.gov, cmgrojean@lbl.gov
\end{center}

\vglue 0.3truecm

\begin{abstract}
\vskip 3pt \noindent
\noindent
\end{abstract}
Brane worlds are theories with extra spatial dimensions in which ordinary
matter is localized on a (3+1) dimensional submanifold. Such theories could
have interesting consequences for particle physics and gravitational 
physics.  In this essay we concentrate on the cosmological constant (CC) 
problem in the context of brane worlds. 
We show how extra-dimensional scenarios may violate Lorentz invariance
in the gravity sector of the effective 4D theory, while particle physics
remains unaffected.
In such theories the usual no-go theorems for adjustment of the 
CC do not apply, and we 
indicate a possible explanation of the smallness of the CC.
Lorentz violating
effects would manifest themselves in gravitational waves travelling with 
a speed different from light, 
which can be searched for in gravitational wave 
experiments. 


\end{titlepage}

\renewcommand{\thefootnote}{(\arabic{footnote})}


\setcounter{equation}{0}
\setcounter{footnote}{0}

It is believed that Einstein's General Relativity (GR) is an inadequate
description of gravity at high
energies because at energies near the Planck scale ($M_{Pl}\sim 10^{19}$ GeV)
the Schwarzschild radius of a system ($G_Nm/c^2$) becomes of the same order as
its Compton length ($\hbar m/c^2$), and the effects of quantum gravity are
important. However, the Planck scale may not necessarily be the scale 
at which modifications to Einstein's theory of gravity first appear. 
In fact, 
contrary to many aspects of particle physics, gravity has not been measured at
distances smaller than about a millimeter.
Therefore,
in principle gravity could begin to  deviate from ordinary 
GR at such scales. One way of modifying gravity is by introducing extra 
dimensions, whose effective size is below the mm scale. However, when 
modifying physics at these relatively low energy scales one has to make 
sure that the standard model of particle physics, which has been 
extremely well tested up to energy scales of the order
of 100 GeV ($\sim 10^{-16}$ cm), is not also modified at those
scales. One way to modify gravity at low energies 
without affecting particle physics is to assume that all the 
fields of the standard model are localized in space to a three-dimensional 
submanifold (``3-brane'') in the higher dimensional world, in which case only 
gravity or other non-standard model fields probe the presence of extra 
dimensions. While this ``brane world''
approach may seem {\it ad hoc}, these principles are
naturally realized in string theory (which is so far the only known
consistent theory of quantum gravity). In fact, string theory necessarily
requires extra dimensions for the consistency of the theory, and gauge
theories localized on branes are a natural part of the theory.

In the context of brane world models several problems in the Standard
Model of particle physics and in gravity have been reformulated, with new
approaches suggested to their solutions.  One of the main motivations for 
considering brane worlds 
is that they suggest new resolutions of the hierarchy problem
(the question of why the energy scale of electroweak interactions is so much 
smaller 
than the scale of gravity).  One way of explaining the hierarchy is by 
assuming that there are flat extra dimensions, which are much larger 
than their natural value $1/M_{Pl}$ \cite{nima}. The presence of such large 
extra
dimensions could lower the {\it fundamental, higher dimensional} Planck
scale all the way to the TeV scale, thereby eliminating the hierarchy
in the fundamental scales of physics. In this scenario 
particle physics experiments with accelerators like the Tevatron or LHC
would directly probe the full theory of quantum gravity, which itself would be
at the TeV scale. Another explanation for the hierarchy between the weak scale
and the Planck scale may be that there is curvature along the 
extra dimensions, causing the natural
scale of the effective 4D theory to depend on the position of the brane along 
the 
extra dimension \cite{RS}.
A typical metric describing such ``warped
spacetimes'' is of the form
\begin{equation}
ds^2=a(y)^2 \eta_{\mu\nu} dx^\mu dx^\nu +dy^2,
\end{equation}
where the warp factor $a(y)$ can induce an exponential hierarchy between
the weak and the Planck scales, and even more strikingly, ensure that 
the brane world observer sees 4D Einstein gravity at larger distances
even without compactification of the extra dimensions.

The main focus of this essay is to explain how extra dimensions
may also help to tackle the cosmological constant problem.
It was first pointed out by Rubakov and Shaposhnikov
\cite{RubSha} that the cosmological constant
problem is reformulated in brane world models.
In ordinary four dimensional cosmology the
contributions to the vacuum energy from gravitational loops 
(${\cal O}(M_{Pl}^4)$), the electroweak phase transition (${\cal O}(10^{-64}
M_{Pl}^4)$) and chiral symmetry breaking (${\cal O}(10^{-76}M_{Pl}^4)$) have
to cancel each other to ${\cal O}(10^{-120}M_{Pl}^4)$ to be consistent
with current bounds on the cosmological constant.
However, in the presence of extra dimensions the four dimensional vacuum 
energy on the brane does not necessarily give rise to
an effective four dimensional cosmological constant. 
Instead the vacuum energy can {\it warp} the
spacetime and introduce a curvature in the bulk
while maintaining a static four dimensional brane world:
in a sense, the energy-induced curvature flows off the brane.
The cosmological constant
problem is then reformulated as the question of why the background warps
in the appropriate fashion without introducing an effective 4D cosmological 
constant; that is, why there would be an exact cancellation between
the brane and the bulk cosmological constants.
{\em Per se} such extra dimensional cancellation mechanisms of the
four dimensional cosmological constant are reminiscent of purely four 
dimensional cancellation mechanisms.  As proposed by Hawking \cite{Hawking}
for example, a four form field strength would provide a contribution to the
cosmological constant whose magnitude is not fixed by the field equations
but appears instead as a constant of integration.  However, the anthropic 
principle must still be invoked in order to explain why that
integration constant happens to be chosen so as to cancel the other 
contributions to the cosmological constant.  Or to say it differently, among
the three classes of maximally symmetric solutions (flat, de Sitter or 
anti-de Sitter) why is the flat solution singled out?  As argued by Weinberg
\cite{Weinberg} all known adjustment mechanisms of the cosmological constant in
a purely four dimensional scenario suffer from similar problems.

In order for extra dimensional scenarios to provide a resolution of the
cosmological constant problem it
would then have to give a positive answer to the following questions:
\begin{enumerate}
\item Can the brane vacuum energy vary continuously in a ``natural'' range
(around the weak scale, for instance) while still maintaining a vanishing
effective 4D cosmological constant?
\item Is there a way to select the flat solution among  maximally symmetric
solutions?
\end{enumerate}

In order for an extra dimensional theory to 
provide a satisfactory resolution to the cosmological constant problem 
it must somehow 
evade Weinberg's no-go theorem for the adjustment of the cosmological
constant. The difficulty is that for an extra dimensional theory to
agree with our observed four dimensional universe, the theory has to have 
an effective 4D description at large distances. However the cosmological 
constant has to be extremely small, at most of ${\cal O}(10^{-3}$ eV$)^4$,
so the CC problem is in some sense a low-energy physics problem.  
Then one should also be able
to understand the cancellation mechanism directly from the effective 4D theory,
and 
one seems to be stuck with Weinberg's no-go theorem.
This is, however, not quite true. 
The freedom of extra dimensional theories
to modify the behavior of gravity in the bulk
while the gauge interactions live on the brane allows 
for the possibility of an effective low-energy description 
which weakly violates 4D Lorentz invariance, without 
contradicting any current observations.  Thus one might hope to 
circumvent the no-go theorem for the adjustment of the cosmological constant
because these theories are fundamentally different than those considered by
Weinberg.
To construct such a theory we
will consider some higher dimensional
geometries in which not only the 4D distance scales vary along the
extra dimensions as in usual warped scenarios,
but also in which the spatial and time scales vary 
in a slightly different way (``asymmetric warping''). A prototypical
example will be given by 5D metrics of the form
\begin{equation}
	\label{metric}
ds^2 = -n^2(r)\, dt^2 + a^2(r)\, \left( \frac{d\sigma^2}{1-k \sigma^2}
+ \sigma^2 d\Omega_2^2 \right) + b^2(r)\, dr^2.
\end{equation}
The coordinate $r$ corresponds to the extra dimension transverse to the brane
and $k=\pm 1, 0$ is the spatial curvature of the 3D sections
parallel to the brane. The induced geometry at the
4D sections of constant $r$ may still be flat, implying that (up to tiny 
quantum
gravitational corrections) particle physics on the brane
will see a Lorentz invariant spacetime. However, different 4D sections
of the metric (\ref{metric}) have a differently defined Lorentz
symmetry: the local speed of light depends
on the position along the extra dimension as
$c(r)=n(r)/a(r)$. Therefore the spacetime (\ref{metric}) globally violates
4D Lorentz invariance, leading to
apparent violations of Lorentz invariance from the brane observer's
point of view due to bulk gravity effects. The important point is that these
effects are restricted to the gravity sector of the effective theory, which
has not been very well measured, while the extremely well measured 
Lorentz invariance of particle physics remains unaffected in these scenarios.

Such asymmetrically warped space-times are actually
quite generic: indeed,
Birkhoff's theorem ensures that the most general solution in the bulk
can be transformed into the black hole metric of the form
(\ref{metric}) with
\begin{equation}
a(r)=r \ \ \mbox{and }\ \
n(r)=\frac{1}{b(r)}=k+\frac{r^2}{l^2}-\frac{\mu}{r^2}+\frac{Q^2}{r^4},
\end{equation}
where $l$ is the radius of curvature of the bulk induced by the 5D vacuum 
energy.
The precise geometry of the black hole depends on the type of sources
and fields which propagate in the bulk; in general, it would be charaterized
by its mass, $\mu$, and its charges, $Q$, which appear
as constants of integration of the equations of motion
and parametrize the asymmetry of the 5D space-time.  The 4D
Lorentz symmetry is restored only when both $\mu$ and $Q$ vanish. These new 
parameters
can now be used to eliminate the fine-tunings that plague the 
symmetrically warped models.

Indeed, it can be shown \cite{us} by solving the Einstein equations around the
brane where standard model fields are localized
(that is, solving the ``Israel junction conditions'')
that the induced metric on a brane
embedded in such a black-hole background
can remain flat whatever the vacuum energy density on the brane.
For instance, a phase transition on the brane might not affect the geometry of 
the brane,
but may instead be compensated for by a change of the mass 
and charge of the BH due
to emission or absorption of gravitational/electromagnetic waves.
Hence, no parameter of the action would have to be tuned to keep the brane 
flat.
This answers the first requirement on solutions of
the cosmological constant problem posed above.
However, one still needs to answer the second question, namely
what selects this flat solution among the 4D maximally symmetric solutions.
The
point is that
other maximally symmetric induced metrics on the brane require
either the mass or the charge of the BH to vanish and so the flat solutions
for which $\mu\not =0$ and $Q\not = 0$ are not continuously connected
to the (anti-)de Sitter solutions. Being an isolated point in
the moduli space, the flat solution is {\it ipso facto} a stable vacuum.
From an effective four dimensional point of view, Weinberg's no-go theorem
is evaded by the Lorentz violating corrections to four dimensional
Einstein gravity.

\begin{figure*}[tb]
\centerline{\includegraphics*[bb=170 2 450 788,angle=-90,width=18cm]{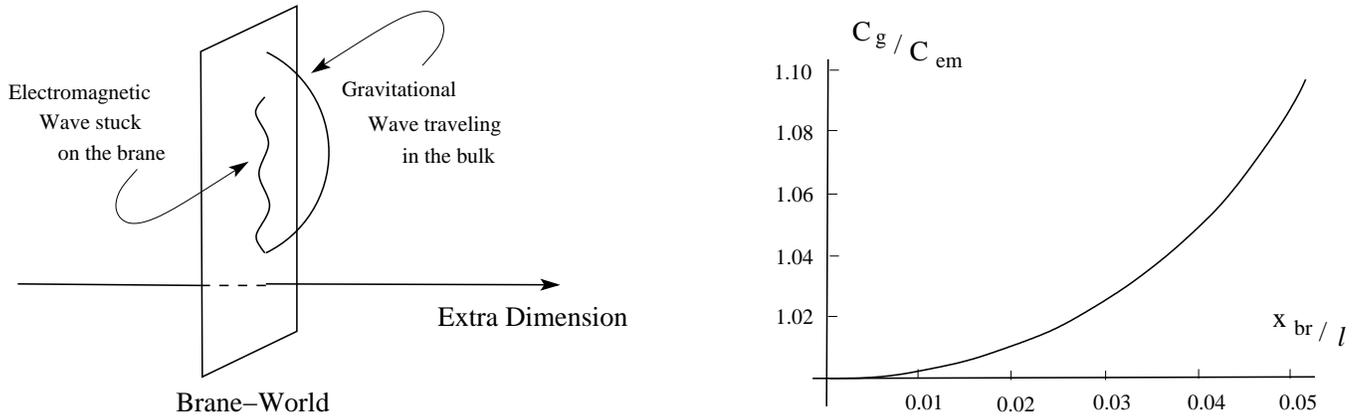}}
\caption[]{A graviton emitted on the brane will travel along
a geodesic in the bulk before returning to the brane.
A photon emitted at the same time can propagate only along the brane and may
wander a shorter distance along the brane than the graviton in the same time.
The 4D effective propagation speed of gravity
is distance dependent ($x_{br}$ is the distance travelled along the brane
and $l$ characterizes the curvature of the bulk).
}
\label{fig:brane}
\end{figure*}

Perhaps the most remarkable property of these asymmetrically warped 
backgrounds is that, in addition to providing a possible
resolution of the CC problem, these backgrounds  have consequences 
that can be verified through gravitational measurements.
%
As we have already stressed, asymmetrically
warped spacetimes break 4D Lorentz invariance in the gravitational sector.
Particle physics will not feel these effects, but gravitational waves
are free to propagate into the bulk and they will necessarily feel the 
effects of the variation of the speed of light along the extra 
dimension. The propagation of gravitational waves in
asymmetrically warped spaces is 
analogous to the propagation of electromagnetic waves through
a medium with a varying index of refraction.  
Gravitational wave propagation
reflects  Fermat's principle, and if the local speed
of light is increasing away from the brane then
gravitational waves propagating between two points on the brane
will take advantage 
by bending slightly into the bulk, 
and will arrive earlier than the electromagnetic waves
which are
stuck to the brane (see Figure). 
Thus in these theories gravitational waves can travel
faster than light! However, 
these faster than light signals do not violate causality with the usual
associated paradoxes.  The apparent violation of causality from the
brane observer's point of view is due to the fact 
that the region of causal contact 
is  actually bigger than one would naively expect from the
ordinary propagation of light in an expanding Universe, but there are no
closed timelike curves in the 5D spacetime that would make the theory 
inconsistent.

The beauty of these models is that 
even extremely  small Lorentz violating effects 
may be measured in gravity wave experiments.  For example, an
astrophysical event such as a distant supernova might generate gravitational
waves which would 
reach future gravity wave detectors before we actually ``see'' the
event. Thus future gravitational
wave experiments like LIGO, VIRGO or LISA might 
discover this unique signature 
of the existence of extra dimensions. If found, 
such evidence may strongly influence
future developments in elementary particle physics, cosmology and
astrophysics.



\end{document}